\documentclass[seceq]{ptptex}
\usepackage[usenames]{color}
\usepackage{graphicx}

\def\lhls{${\cal L}_h/{\cal L}_s$}

\newbox\grsign \setbox\grsign=\hbox{$>$} \newdimen\grdimen %
\grdimen=\ht\grsign
\newbox\simlessbox \newbox\simgreatbox \newbox\simpropbox
\setbox\simgreatbox=\hbox{\raise.5ex\hbox{$>$}\llap
     {\lower.5ex\hbox{$\sim$}}}\ht1=\grdimen\dp1=0pt
\setbox\simlessbox=\hbox{\raise.5ex\hbox{$<$}\llap
     {\lower.5ex\hbox{$\sim$}}}\ht2=\grdimen\dp2=0pt
\setbox\simpropbox=\hbox{\raise.5ex\hbox{$\propto$}\llap
     {\lower.5ex\hbox{$\sim$}}}\ht2=\grdimen\dp2=0pt
\def\simgreat{\mathrel{\copy\simgreatbox}}
\def\simless{\mathrel{\copy\simlessbox}}

\title{
Spectral transitions in X-ray Binaries
}

\author{
Chris {\sc Done} and Marek {\sc Gierli\'nski}%
}

\inst{
Department of Physics, University of Durham, South Road, Durham DH1 3LE, UK
}

\abst{We systematically analyzed about a terabyte of data from Galactic
binary systems to characterize how these spectra change as a function
of mass accretion rate.  With this volume of data we are able to
identify the differences in behaviour of black holes and neutron
stars, implying that black holes really do have an event horizon. The
spectral evolution seen as a function of $L/L_{Edd}$ can be linked to
current state-of-the-art numerical simulations of accretion flows, 
and we develop qualitative and quantitative models of the
spectral changes in all types of binary systems. The key ingredient of
this picture is that the major hard-soft spectral transition is driven
by a changing inner radius of the accretion disc. This also
qualitatively explains the changes seen in the variability power
spectra of these objects. This lays the foundation for a
self-consistent, theoretically motivated framework in which to explain
both the spectral and timing properties of accreting compact objects.}

\begin{document}

\maketitle

\section{Introduction}

One of the key puzzles in X-ray astronomy is to understand accretion
flows in a strong gravitational field. This applies to both Active
Galactic Nuclei (AGN) and Quasars, where the accretion is onto a
supermassive black hole, the puzzling ultraluminous compact sources
which may be intermediate mass black holes, the stellar mass Galactic
black holes (GBH) and even the neutron star systems.  Neutron star
radii are of order three Schwarzchild radii, i.e. the same as that for
the last stable orbit of material around a black hole. Thus they have
very similar gravitational potentials so should have very similar
accretion flows, though of course with the major difference that
neutron stars have a solid surface, so can have a boundary layer
and a stellar magnetic field.

The main premise is that progress in understanding accretion in {\em
any} of these objects should give us some pointers to understanding
accretion in {\em all} of them. Galactic sources are intrinsically
less luminous, but a great deal closer than the AGN, so generally are
much brighter. Galactic black holes are also often generically
transient, showing large variability on timescales from milliseconds
to years.  These give us a sequence of spectra at differing mass
accretion rates onto the central object, allowing us to test accretion
models. 

\section{Accretion Flows in Galactic Black Holes}

Black holes are very simple objects, possessing only mass and spin.
In steady state the appearance of the accretion flow should be
completely determined by these parameters, together with the mass
accretion rate and some weak dependence on inclination angle.
Much of the dependence on mass can be removed by scaling the accretion
rate to the Eddington accretion rate, as sources emitting at similar
fractions of the Eddington luminosity, $L/L_{Edd}$, should have
similar accretion flows.

The best known solution for the accretion flow equations is the
Shakura--Sunyaev optically thick disc\cite{ss73}, hereafter SS.  Viscous
stresses convert some of the gravitational potential energy to heat,
producing a rather robust, quasi-blackbody spectrum, with temperature
of $kT_{disc} \sim 1$~keV for accretion rates around Eddington onto a
$10M_\odot$ GBH. Such spectra are seen, but are generally accompanied
by a weak (ultrasoft state: US), moderate (high state: HS) or strong
(very high state and intermediate state: VHS) X-ray tail to higher
energies. Together these form the soft states, which are seen at high
$L/L_{Edd}$. However, at low $L/L_{Edd}$ these objects can also show
spectra which look entirely unlike a disc, peaking instead at $\sim
100$ keV (low/hard states: LS). Fig. 1a shows representative spectra
from all these GBH states\cite{tl95}.

To produce any emission at energies substantially higher than that of
the disc {\em requires} that some fraction of the gravitational energy
is dissipated in regions which are optically thin, so that a few
electrons gain a large fraction of the energy. These energetic
electrons can produce hard X-rays by Compton upscattering lower energy
photons, and the shape of this spectrum is determined by the ratio of
power in the hot electrons to that in the seed photons illuminating
them, \lhls. 

While such Comptonization models can explain the broad band spectral
shapes, they do not address the underlying problem of the {\em
physical origin} of the hot electrons, or indeed the range \lhls\
required to produce the very different spectra shown in Fig. 1a.  We
can get some insight into these more fundamental issues from recent
advances in understanding the physical nature of the accretion disc
viscosity as a magnetic dynamo\cite{bh91}.  Numerical
simulations show that any seed magnetic field can be continuously
amplified by the differential rotation of the disc material, and
dissipated through reconnection events. Including radiative cooling
gives an accretion disc structure which bears some resemblance to the
standard accretion disc models, but with some of the magnetic
reconnection occurring above the disc as magnetic field loops buoyantly
rise to the surface, reconnecting above the bulk of the material in
an optically thin environment\cite{t04}\cite{ms00}.

However, these physical viscosity simulations also show that an alternative,
{\em non-disc} solution exists where the whole accretion flow is
optically thin, so cannot efficiently cool. The accretion flow
forms a hot, 
geometrically thick structure, qualitatively similar to the Advection
Dominated Accretion Flows\cite{ny95}, but considerably more
complex in detail, with convection and outflow as well as advection\cite{hb02}.

The existence of two very different accretion flow structures gives a
very natural explanation for the two very different types of spectra
seen from the GBH.  At low $L/L_{Edd}$ the inner optically thick disc
could be replaced by an optically thin flow. There are few photons from the
disc which illuminate the flow, so \lhls$\gg1$ and the Comptonized
spectra are hard. When the mass accretion rate increases, the flow
becomes optically thick, and collapses into an SS disc. The dramatic
increase in disc flux drives the hard-soft state transition\cite{e97}.
A weak tail on the dominant disc emission can be produced by
occasional magnetic field loops buoyantly rising to the surface,
reconnecting above the bulk of the material in an optically thin
environment (US).  Increasing the ratio of power dissipated above the
surface to that in the disc increases \lhls, increasing the importance
of the hard X-ray tail. However, the {\em geometry} of the soft states
sets a limit to \lhls. Flares {\em above} a disc illuminate the disc
surface, where some fraction are absorbed and thermalized. This adds
to the intrinsic disc emission, fixing \lhls$\simless 1$ in the limit
where the flares cover most of the disc surface\cite{hm93},
which always results in a soft Comptonized spectrum, forming a
power law with energy index $\alpha \simgreat 1$ (VHS). Fig. 2
illustrates the geometries inferred for each state.

\section{Spectral evolution as a function of luminosity for Black Holes}

The previous section outlined a physically motivated model in which
spectral changes (particularly the hard-soft spectral transition) are
driven by a changing geometry. There now exists an enormous amount of
data from the X-ray binary systems which can be used to test this. 
We looked at of order a terabyte of data from the {\it RXTE} satellite
to systematically analyze all the available spectra from many black
hole systems. Clearly, doing a detailed spectral analysis
of so much data is impractical. Broad band 'colours' have long been
used in neutron star X-ray binaries to get an overview of source
behaviour. The problem is that these colours are  often defined
using counts within a certain energy range, so are generally dependent
on the instrument
response, as well as on the absorbing
column toward the source. We want to compare many different black
holes, so instead use {\em intrinsic} colour, i.e. unabsorbed flux ratios
over a given energy band. To do this we need a physical model. Plainly
there can be emission from an accretion disc, together with a higher
energy component from Comptonization. Reflection of this emission from
the surface of the accretion disc can also contribute to the
spectrum. Thus we use a model consisting of a multicolour accretion
disc, Comptonized emission (which is {\em not} a power law at energies
close to either the seed photon temperature or the mean electron
energy), with Gaussian line and smeared edge to roughly model the
reflected spectral features, together with galactic absorption.

We use this model to fit the RXTE PCA data from all available black
holes which are not strongly affected by absorption $N_H< 2\times
10^{22}$ cm$^{-2}$. We choose 4 energy bands, 3-4 keV, 4-6.4 keV,
6.4-9.7 keV and 9.7-16 keV, and integrate the unabsorbed model over
these ranges to form {\em intrinsic} colours, which roughly describe
the {\em intrinsic} spectral slope (corrected for absorption and
instrument response by model fitting) below and above 6~keV,
respectively. We use the (generally) fairly well known distance to
convert the extrapolated bolometric flux to total luminosity. Again,
since the mass of the central object is fairly well known we can
translate the bolometric luminosity into a fraction of the Eddington
luminosity.

Fig. 1b shows {\em all} the data from many different black holes as a
colour-colour diagram.  {\em All} the black holes are consistent with the
{\em same} spectral evolution as a function of increasing luminosity
$L$ (scaled in terms of the Eddington luminosity, $L_{Edd}$) for
$L/L_{Edd}\sim$ 0.001--0.5. At low luminosity the spectra are hardest
(with largest values of both soft and hard colours). As the luminosity
increases the spectrum softens, forming a well 
defined diagonal track in the colour-colour plots. This expands into
a much wider range colours for the soft states, where there is an
amazing variety of spectral shapes for the same $L/L_{\rm Edd}$
\cite{dg03}.

\begin{figure}[!t]
  \leavevmode
  \begin{center}
  \begin{tabular}{c}
   \includegraphics[width=0.9\textwidth]{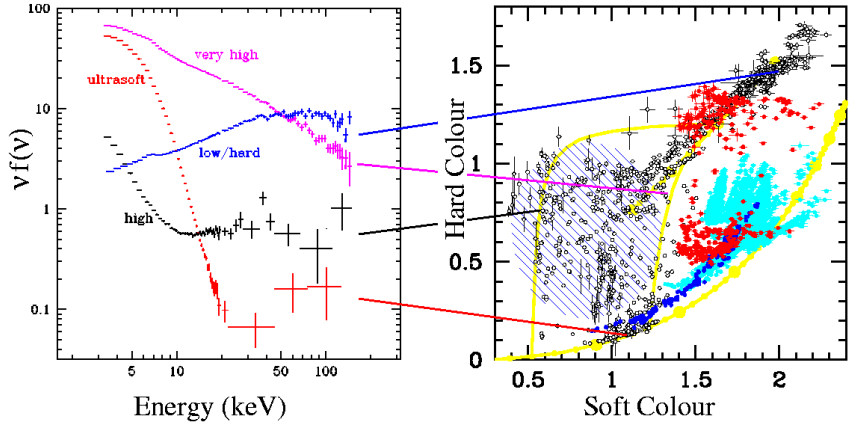}
  \end{tabular}
  \caption{(a) illustrates the range of spectral shapes seen from the GBH. (b)
compresses all the spectral information into intrinsic colours for
black holes (black), 
atoll systems (red points), Z sources (cyan points) and the odd
Z source Cir X--1 (blue points). The blue shaded area indicates the
section of the colour-colour diagram corresponding to the high/soft
state spectra (Fig. 1a) which is only occupied by black holes\cite{dg03}.
}
  \end{center}
\end{figure}

\begin{figure}[!b]
  \leavevmode
  \begin{center}
  \begin{tabular}{c}
  \includegraphics[width=0.9\textwidth]{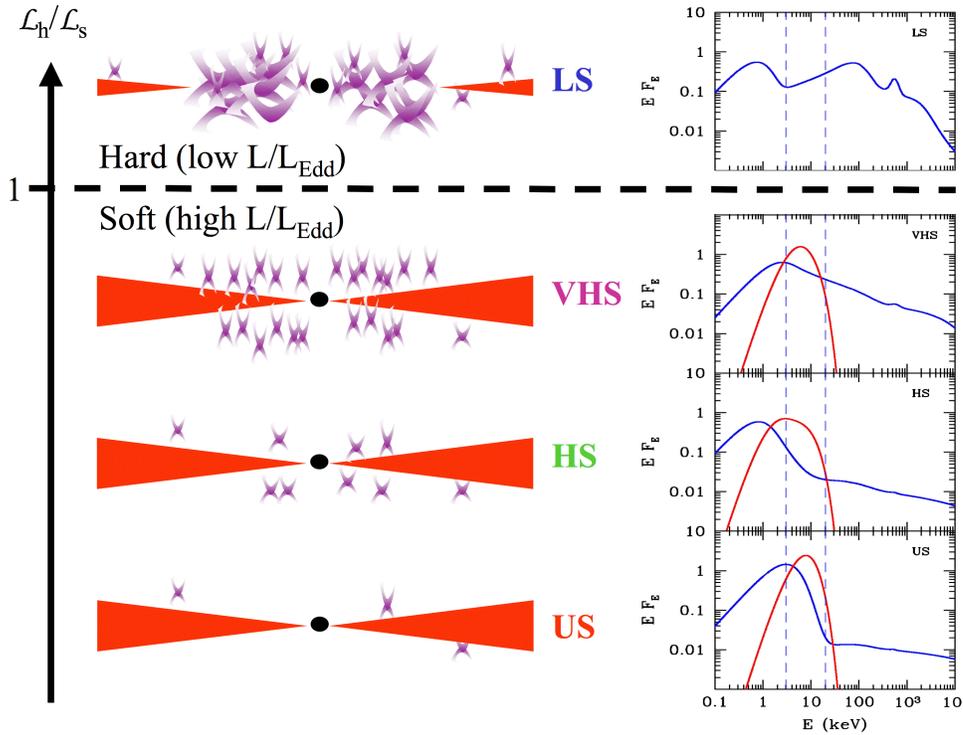} 
  \end{tabular}
  \caption{Sketched geometries corresponding to the spectral states as
    a function of \lhls, together with the {\sc eqpair} spectra for
    GBH (blue). At high $L/L_{Edd}$ the identical accretion
    flow in atoll systems has an additional component from an
    optically thick boundary layer (red).}
  \label{fig:geom}
  \end{center}
\end{figure}

\section{Accretion flows in neutron star binaries}

Neutron stars without a strong magnetic field ($B<10^{12}$ G) come in
two flavors, named atolls and Z sources. Z sources are named after a
Z-shaped track they produce on an X-ray colour-colour diagram, while
atolls are named after their C (or atoll) shaped track.  These
differences between the two Low Mass X-ray Binary categories probably
reflect differences in both mass accretion rate and magnetic field,
$B$, with the Z sources having high luminosity (typically more than 50
per cent of the Eddington limit) and magnetic field ($B \ge 10^9$ G)
while the atolls have lower luminosity (generally less than 10 per
cent of Eddington) and low magnetic field ($B \ll 10^9$ G)
\cite{hv89}.

We repeat the intrinsic colour analysis for 4 different transient
atoll systems spanning $L/L_{Edd}\sim$ 0.001--0.5. These are plotted
as the red points in Fig. 1b, and are all consistent with the same
spectral evolution, forming a large, Z shaped track with increasing
$L/L_{\rm Edd}$. The spectra at low luminosities are hard, resembling
the black hole low/hard spectra\cite{bv94}, then as
the luminosity increases the colours move horizontally to the right,
then there is an abrupt transition, where both hard and soft colour
decrease, forming the start of the lower branch of the Z (termed the
banana branch). This hard-soft transition is plainly reminiscent of
the black hole behaviour, but the atolls evolve very differently with
$L/L_{Edd}$ on the colour-colour diagram.

Ironically, the atolls actually form a much more convincing Z shaped
track on the colour-colour diagrams than the Z sources! The newly
discovered upper branch of the Z in atolls is only seen at low
luminosities, and transforms their C (or atoll) shape into a Z\cite{gd02}\cite{m02}. The Z sources (cyan
points in Fig. 1b) form a much less coherent picture, with different Z
sources having different colours rather than forming a single track
despite all being at similar luminosities of $L/L_{Edd}\sim 1-2$,
except for the odd source Cir X--1 (blue points) which spans
$L/L_{Edd}\sim 2-10$.

Since we have calculated intrinsic colours, then not only can we
compare many objects in a given class, we can also put the black holes
and neutron star systems on the same diagram (Fig. 1b). The
differences in the black hole and atoll tracks are then obvious, as is
the fact that there are parts of the colour-colour diagram which are
{\em only} occupied by black holes. This is not simply that the source
shows soft colours, as it has long been known that Cir X-1 can show
colours similar to those of the ultrasoft black hole spectra. However,
{\em no} type of disc acceting neutron star even has colours similar
to those of the classic high/soft spectra (Fig. 1a), where there is
low temperature thermal emission from the disc together with a power
law tail to much higher energies.  The obvious interpretation is that
there are physical, observable spectral differences due to the
presence/absence of a solid surface\cite{dg03}.

\section{Detailed modeling of the accretion flows in GBH and atolls}

The colour-colour data shown in Fig. 1b can be qualitatively and
quantitatively tied into the theoretical models of accretion flows
discussed in \S 2. Taking the black holes first,
if the inner disc is replaced by a
hot flow at low $L/L_{Edd}$ then there are few seed photons from the
disc so the spectrum is hard. As the mass accretion rate increases,
the truncation radius of the disc decreases, so it penetrates further
into the hot flow. A larger fraction of the disc photons are
intercepted by the hot flow, increasing the Compton cooling. This
steepens the Comptonized spectrum, but since this dominates the whole
X-ray band then the whole spectrum softens, giving rise to the
diagonal track on the colour-colour diagram.  When the mass accretion
rate is such that the hot flow starts to become optically thick then
it collapses into a standard SS disc, abruptly softening
the spectrum.

This schematic picture (see Fig. 2) can be translated into a {\em
quantitative} model using the sophisticated Comptonization code, {\sc
eqpair}\cite{c99}.  The key advantage of this code is that it does
not assume the steady state electron distribution, rather it {\em
calculates} it by balancing heating (injection of power ${\cal L}_h$
into thermal and/or non-thermal electrons) and cooling processes
(Compton cooling, which depends on ${\cal L}_s$, and Coulomb
collisions). The resulting spectrum depends primarily on \lhls,
i.e. on the geometry, and can have complex curvature in the
Comptonized emission. This non-power law Comptonized continuum is
required in order to fit broad bandpass individual
spectra from all states (HS\cite{g99}, VHS\cite{z01,gd03},
LS\cite{z02}).

Fig. 3a shows a grid of colours resulting from the {\sc eqpair} code
for \lhls changing from 30 (top right of the diagonal branch) -- 0.01
(softest hard colours), assuming seed photons from the disc at 0.3 --
1.2 keV as expected for the observed range in $L/L_{Edd}$ for a
standard disc.  Changing only these two {\em physical} parameters can
describe {\em all} the spectra seen from the GBH. These model spectra
for each state are shown by the blue lines in the right hand panel of
Fig. 2, with the blue dotted lines showing the energy range of the PCA
data over which the colours are measured\cite{dg03}.

\begin{figure}[!t]
  \leavevmode
  \begin{center}
  \begin{tabular}{c}
  \includegraphics[width=0.33\textwidth]{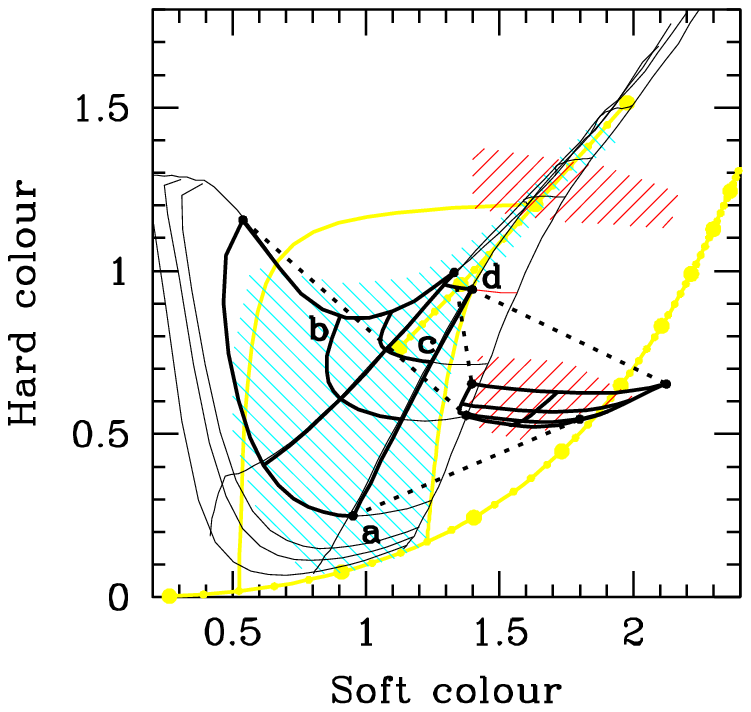} 
  \includegraphics[width=0.6\textwidth]{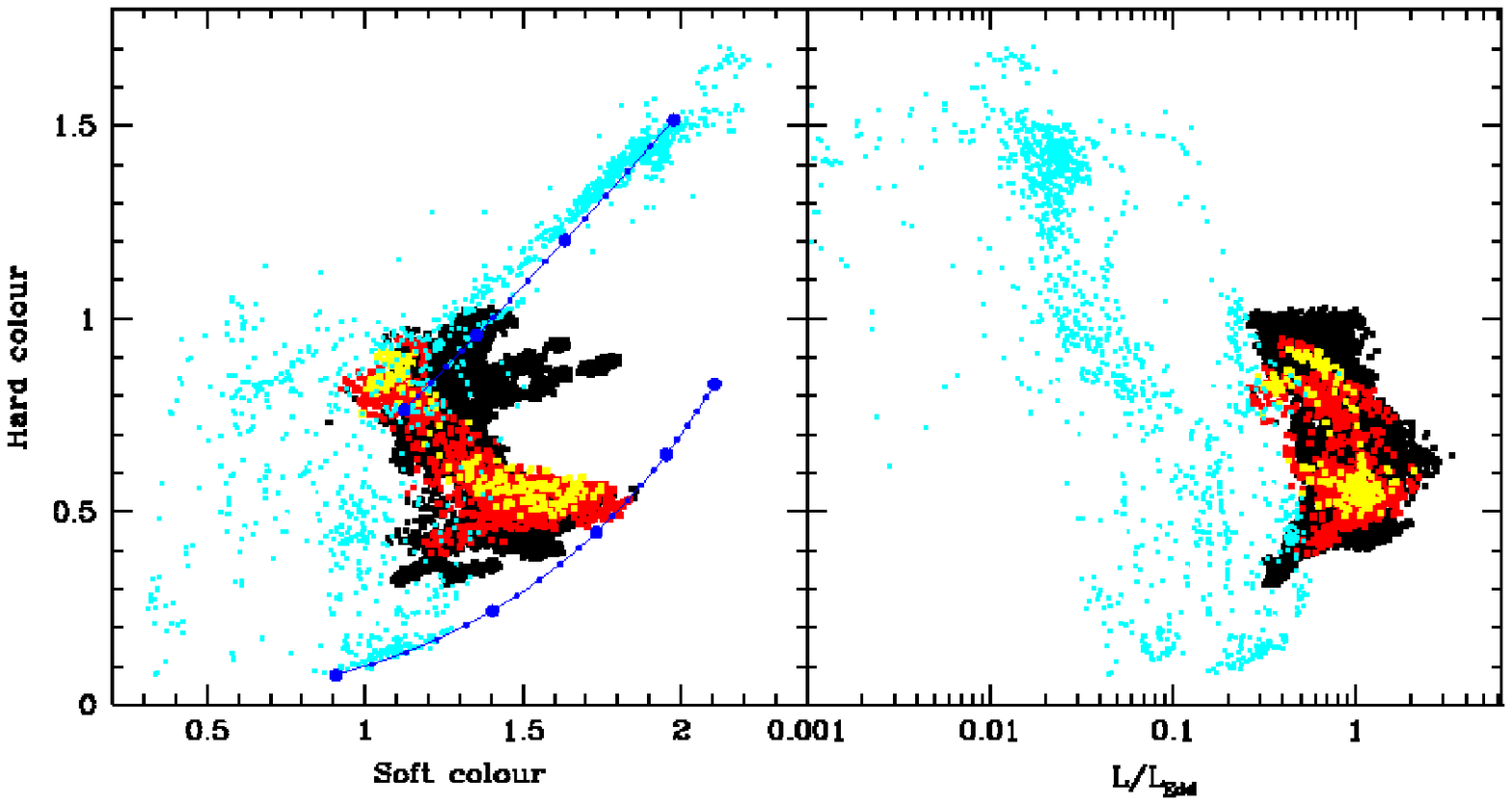}
  \end{tabular}
  \caption{(a) shows the grid of {\sc eqpair} models for the black holes.
The cyan and red shaded areas
show the region of the colour-colour diagram
occupied by data from the black holes and atolls, respectively.
The addition of an optically thick boundary layer
onto these model spectra is enough to transform all the soft
state black hole colours onto the banana branch\cite{dg03} (Fig. 2). (b) shows
the colour-colour and colour-luminosity diagram for GRS~1915+105, with
red and yellow points indicating data taken during times
of the characteristic limit cycle variability, while the black points
show times where the variability is similar to that seen from other
black holes. The cyan points show data from 
all the other black holes for comparison\cite{dwg04}.
}
  \end{center}
\end{figure}

The {\em same} models can also describe the disc accreting neutron
stars but with the difference that there is a solid surface/boundary
layer.  Schematically, at low luminosity, the disc is truncated a long
way from the neutron star, and the boundary layer is mostly optically
thin, so it joins smoothly onto the emission from the inner accretion
flow.  Reprocessed photons from the X-ray illuminated surface form the
seed photons for Compton cooling of the inner flow/boundary layer
emission, so the PCA spectrum is dominated by this Comptonized
component.  As the mass accretion rate increases, the disc starts to
move inward, but the cooling is dominated by seed photons from the
neutron star rather than from the disc, so the geometry and hence the
high energy spectral shape does not change. The atolls keep constant
hard colour, but the soft colour increases as the seed photon energy
moves into the PCA band. Eventually the mass accretion rate becomes
high enough to make the inner flow/boundary layer become optically
thick. This triggers the collapse of the inner flow into a standard
disc, softening the spectrum at low energies, abruptly decreasing the
soft colour. The boundary layer properties also change as it is now
optically thick, so its cooling is dominated by quasi-blackbody rather
than Compton processes. Since blackbody cooling is much more
effective, the temperature of the boundary layer drops, softening the
hard spectrum and decreasing its colour.  Thus as the inner
flow/boundary layer becomes optically thick this causes {\em both} the
inner flow to change to a disc {\em and} the boundary layer emission
to thermalize. This reduces both the hard and soft colours, so the
track moves abruptly down and to the left during this transition.
After this then increasing the mass accretion rate increases the disc
temperature, so the track moves to higher soft
colour\cite{gd02}\cite{dg03}.  The 'banana branch' is then
analogous to the high/soft state in the galactic black holes, but with
additional luminosity from the boundary layer.

We can look at the quantitative predictions of the model for the high
mass accretion rate atolls by simply adding optically thick emission
from a boundary layer to the previous black hole {\sc eqpair} models,
as shown by the red lines on the spectra in Fig. 2. With this
addition, the {\em same} {\sc eqpair} model grid which gave a whole variety
of soft state spectra for the black holes collapses into a well
defined 'banana branch' for the atoll systems (Fig. 3a)\cite{dg03}.

The similarity of topology on the colour-colour diagram of the atolls
and Z sources again supports the truncated disc models. It has long
been suggested that the jump between the upper and lower tracks of the
Z in Z sources is formed by the interaction of the disc and magnetic
field\cite{l91}. At the lower mass accretion rates the disc is
truncated by the magnetic field, but as $L/L_{Edd}$ increases, the
increasing ram pressure of the disc material finally overcomes the
magnetic pressure, allowing the disc to penetrate to the neutron star
surface.  The high mass accretion rates mean that the inner
flow/boundary layer is always optically thick, so these systems never
show the hard spectra seen from upper branch of the Z in atolls. There
{\em are} real physical differences between Z sources and
atolls\cite{r04}. The topology of their tracks on the colour-colour
diagram are both driven by the same changing geometry, but the
mechanism causing the disc truncation is different.

\section{GRS~1915+105}

Even the pathologically strange source GRS~1915+105 can be fit into
this picture. This black hole shows a unique variability behaviour in
about 50\% of the XTE data collected from this source, where the disc
continually switches between being hot and bright with small inner
disc radius to cooler and dimmer with much larger inner disc
radius. This can be interpreted as resulting from a limit-cycle
instability in the inner accretion disc, such that it is continually
emptying and refilling\cite{b97}\cite{b00}.

Fig. 3b and c shows the colours and inferred luminosity (assuming a
distance of 12.5~kpc) for all the GRS1915+105 data in the XTE
database. We fit each 128s spectra as before, including careful
modeling of the heavy absorption\cite{dwg04}.
Spectra taken from periods where the limit cycle variability is
present are coloured red (most extreme, rms variability $\ge$ 60~\%)
and yellow (rms variability between 40--60~\%). The cyan points mark
the data from all the other 'normal' black holes.  It is plain that
the instability occurs in only the most luminous spectra, those which
have $L > L_{\rm Edd}$. The luminosity never drops below $\sim$0.3
$L_{\rm Edd}$, explaining why it never goes into the hard state
(typically seen only at luminosities lower than a few percent of
Eddington). It also explains why it {\em never\/} goes into the `black
hole only' area of the colour-colour diagram.  Spectra in this region,
characterized by a low-temperature disc component together with a weak
hard tail, are seen predominantly around $\sim$0.1 $L_{\rm Edd}$. The
persistently high luminosity of GRS1915+105 takes it above these low
disc temperatures. 

However, superEddington luminosity seems to be only a necessary, not
sufficient condition for the disc instability. There are several low
variability spectra from GRS1915+105 which exceed the Eddington limit,
and there are several of the other black holes which reach comparable
luminosities, but do not show the limit cycle. The colour-luminosity
diagram for the 'normal' black holes shows clearly that there is not a
one-to-one relation between luminosity and spectral state. It seems
that the spectral evolution is driven by the average mass
accretion rate, {\em not} the instantaneous mass accretion rate
as inferred
from the X-ray luminosity. There is some much longer timescale in the
system, presumably tied to the response of the disc and/or inner flow
\cite{v00,v01}. 

GRS1915+105 is not in a separate class from `normal' black holes. When
it is at the same $L/L_{\rm Edd}$ then its spectra and (more
importantly) time variability behaviour are similar to that seen in
the 'normal' black holes. Its unique limit cycle variability is
probably linked to the fact that it is the {\em only} black hole to
radiate at high (super Eddington) luminosities for a sustained period
of time.

\section{Reflection and QPO's}

A truncated disc/inner hot flow model can explain the spectral
evolution of all the different classes of low mass X-ray binaries. It
is also consistent with results from detailed spectral fitting of
individual, broad bandpass spectra. One of the best ways to track the
disc is to look at the reflected line and continuum which results from
X-ray illumination of the optically thick material. The amount of
reflection and line show the solid angle subtended by the optically
thick material, while the strength of relativistic smearing by special
and general relativistic effects shows the extent to which the disc
penetrates into the strong gravitational field\cite{f00}.

All detailed fits to the low/hard state spectra in black holes are
consistent with a rather smaller amount of reflection and relativistic
smearing than would be expected for a disc which extends down to the
last stable orbit at 3 Schwarzchild radii\cite{gcr99,zls99,bdn03} More
reflection and smearing are seen in high state
spectra\cite{g99,gcr99,m04}.

Models with a moving inner disc radius at low mass accretion rates
also can qualitatively explain the variability power spectra of these
sources, which show characteristic frequencies in the form both of
breaks and Quasi Periodic Oscillations (QPO's). These features are
related ($f_{break}\sim 10 f_{QPO}$ for the low frequency QPO), and
they {\em move}, with the frequencies generally being higher
(indicating smaller size scales) at higher $\dot{m}$\cite{v00}.
Recent progress has concentrated on the
similarity between the relationship between the QPO and break
frequencies in black holes {\em and} neutron star systems\cite{v00}.
If they truly are the same phenomena then
the mechanism {\em must} be connected to the accretion disc properties
and not to the magnetosphere or surface of the neutron star. While the
variability is not yet understood in detail, {\em all} QPO and break
frequency models use a sharp transition in the accretion disc in some
form to pick out a preferred timescale\cite{v00}, so by
far the easiest way to change these frequencies is to change the inner
disc radius\cite{cgr01}.

\section{Conclusions}

We can form a unified picture of the spectral evolution of all types
of low mass X-ray binaries, where the major hard-soft spectral
transition is driven by a changing inner disc radius linked to the
collapse of an optically thin, hot inner flow. We see clear
differences between the black holes and disc accreting neutron stars,
both in the form of a unique black hole spectral signature (the
high/soft state) and in terms of the evolution of their spectral shape
with $L/L_{Edd}$. These differences can be modeled qualitatively and
quantitatively as the {\em same} accretion flow onto a {\em different}
object: neutron stars have a surface so have a boundary layer, while
black holes have an event horizon!

\smallskip

\end{document}